# A note on proper curvature collineations in Bianchi types $VI_0$ and $VII_0$ space-times

Ghulam Shabbir and Amjad Ali

Faculty of Engineering Sciences,

GIK Institute of Engineering Sciences and Technology,

Topi, Swabi, NWFP, Pakistan.

Email: shabbir@giki.edu.pk

**Abstract**

We study proper curvature collineations in the most general form of the Bianchi types $VI_0$ and $VII_0$ space-times using the rank of the $6 \times 6$ Riemann matrix and direct integration technique. It is shown that when the above space-times admit proper curvature collineations, they form an infinite dimensional vector space.

## 1. INTRODUCTION

The aim of this paper is to find the existence of proper curvature collineations in the Bianchi types $VI_0$ and $VII_0$ space-times. Since the curvature tensor is an important in differential geometry and general relativity. Hence the study of its symmetries is important. Different approaches [1-22] were adopted to study curvature collineations. Throughout $M$ represents a four dimensional, connected, Hausdorff space-time manifold with Lorentz metric $g$ of signature (-, +, +, +). The curvature tensor associated with $g_{ab}$, through the Levi-Civita connection, is denoted in component form by $R^a{}_{bcd}$. The usual covariant, partial and Lie derivatives are denoted by a semicolon, a comma and the symbol $L$, respectively. Round and square brackets denote the usual symmetrization and skew-symmetrization, respectively. Here, $M$ is assumed non-flat in the sense that the curvature tensor does not vanish over any non-empty open subset of $M$.

Any vector field $X$ on $M$ can be decomposed as

$$X_{a;b} = \frac{1}{2} h_{ab} + G_{ab} \tag{1}$$

where $h_{ab}(=h_{ba})=L_X g_{ab}$ is a symmetric and $G_{ab}(=-G_{ba})$ is a skew symmetric tensor on $M$. If $h_{ab;c}=0$, $X$ is said to be *affine* and further satisfies $h_{ab}=2cg_{ab}, c \in R$ then $X$ is said to be *homothetic* (and *Killing* if c = 0). The vector field $X$ is said to be proper affine if it is not homothetic vector field and also $X$ is said to be proper homothetic vector field if it is not Killing vector field.

A vector field $X$ on $M$ is called a curvature collineation (CC) if it satisfies [1]

$$L_X R^a{}_{bcd} = 0 \tag{2}$$

or equivalently,

$$R^a{}_{bcd;e} X^e + R^a{}_{ecd} X^e{}_{;b} + R^a{}_{bed} X^e{}_{;c} + R^a{}_{bce} X^e{}_{;d} - R^e{}_{bcd} X^a{}_{;e} = 0.$$

The vector field $X$ is said to be proper CC if it is not affine [4] on $M$.

## 2. CLASSIFICATION OF THE RIEMANN TENSORS

In this section we will classify the Riemann tensor in terms of its rank and bivector decomposition.

The rank of the Riemann tensor is the rank of the $6 \times 6$ symmetric matrix derived in a well known way [4]. The rank of the Riemann tensor at $p$ is the rank of the linear map $f$ which maps the vector space of all bivectors $G$ at $p$ to itself and is defined by $f: G^{ab} \rightarrow R^{ab}{}_{cd} G^{cd}$. Define the subspace $N_p$ of the tangent space $T_p M$ consisting of those members $k$ of $T_p M$ which satisfy the relation

$$R_{abcd} k^d = 0 \tag{3}$$

Then the Riemann tensor at $p$ satisfies exactly one of the following algebraic conditions [4].

**Class B**

The rank is 2 and the range of $f$ is spanned by the dual pair of non-null simple bivectors and dim $N_p = 0$. The Riemann tensor at $p$ takes the form

$$R_{abcd} = \alpha\, G_{ab} G_{cd} + \beta\, \overset{*}{G}_{ab} \overset{*}{G}_{cd} \tag{4}$$

where $G$ and its dual $\overset{*}{G}$ are the (unique up to scaling) simple non-null spacelike and timelike bivectors in the range of $f$, respectively and $\alpha, \beta \in R$.

**Class C**

The rank is 2 or 3 and there exists a unique (up to scaling) solution say, $k$ of (3) (and so $\dim N_p = 1$). The Riemann tensor at $p$ takes the form

$$R_{abcd} = \sum_{i,j=1}^{3} \alpha_{ij} G^i{}_{ab} G^j{}_{cd} \tag{5}$$

where $\alpha_{ij} \in R$ for all $i,j$ and $G^i{}_{ab} k^b = 0$ for each of the bivectors $G^i$ which span the range of $f$.

**Class D**

Here the rank of the curvature matrix is 1. The range of the map $f$ is spanned by a single bivector $G$, say, which has to be simple because the symmetry of Riemann tensor $R_{a[bcd]} = 0$ means $G_{a[b} G_{cd]} = 0$. Then it follows from a standard result that $G$ is simple. The curvature tensor admits exactly two independent solutions $k, u$ of equation (3) so that $\dim N_p = 2$. The Riemann tensor at $p$ takes the form

$$R_{abcd} = \alpha\, G_{ab} G_{cd}, \tag{6}$$

where $\alpha \in R$ and $G$ is simple bivector with blade orthogonal to $k$ and $u$.

**Class O**

The rank of the curvature matrix is 0 (so that $R_{abcd} = 0$) and $\dim N_p = 4$.

**Class A**

The Riemann tensor is said to be of class A at $p$ if it is not of class B, C, D or O. Here always $\dim N_p = 0$.

A study of the CCS for the classes A, B, C, D and O can be found in [4, 5].

**4. Main Results**

Consider Bianchi types $VI_0$ and $VII_0$ space-times in usual coordinate system $(t,x,y,z)$ (labeled by $(x^0, x^1, x^2, x^3)$, respectively) with line element [23]

$$ds^2 = -dt^2 + \left(A f^2(z) + B h^2(z)\right) dx^2 + \left(A h^2(z) + B f^2(z)\right) dy^2 + 2(A+B) f(z) h(z) dx\, dy + C(t) dz^2, \tag{7}$$

where $A$, $B$ and $C$ are nowhere zero functions of $t$ only. For $f(z) = \cosh z$, $h(z) = \sinh z$ or $f(z) = \cos z$, $h(z) = \sin z$ the above space-time (7) becomes Bianchi type $VI_0$ or $VII_0$, respectively. The above space-time admits three linearly independent Killing vector fields which are

$$\frac{\partial}{\partial x}, \frac{\partial}{\partial y}, -y\frac{\partial}{\partial x} - x\frac{\partial}{\partial y} + \frac{\partial}{\partial z}. \tag{8}$$

The non-zero components of the Riemann tensor are [22]

$$R_{0101} = -\frac{1}{4}\left[\left(\frac{2A\ddot{A} - \dot{A}^2}{A}\right)f^2(z) + \left(\frac{2B\ddot{B} - \dot{B}^2}{B}\right)h^2(z)\right] = \alpha_1,$$

$$R_{0102} = -\frac{1}{4}\left[\left(\frac{2A\ddot{A} - \dot{A}^2}{A}\right) + \left(\frac{2B\ddot{B} - \dot{B}^2}{B}\right)\right]f(z)h(z) = \alpha_7,$$

$$R_{0113} = R_{0223} = -\frac{1}{4ABC}\left[A^2\dot{B}C + \dot{A}B^2C + 2AB\dot{C}(A+B) - 3ABC(\dot{A}+\dot{B})\right]f(z)h(z) = \alpha_8,$$

$$R_{0123} = -\frac{1}{4}\left[\left\{(A+B)\left(\frac{\dot{A}}{A} + \frac{\dot{C}}{C}\right) - 2(\dot{A}+\dot{B})\right\}h^2(z) + \left\{(A+B)\left(\frac{\dot{B}}{B} + \frac{\dot{C}}{C}\right) - 2(\dot{A}+\dot{B})\right\}f^2(z)\right]$$

$$= \alpha_9, \quad R_{0202} = -\frac{1}{4}\left[\left(\frac{2A\ddot{A} - \dot{A}^2}{A}\right)h^2(z) + \left(\frac{2B\ddot{B} - \dot{B}^2}{B}\right)f^2(z)\right] = \alpha_2,$$

$$R_{0213} = -\frac{1}{4}\left[\left\{(A+B)\left(\frac{\dot{A}}{A} + \frac{\dot{C}}{C}\right) - 2(\dot{A}+\dot{B})\right\}f^2(z) + \left\{(A+B)\left(\frac{\dot{B}}{B} + \frac{\dot{C}}{C}\right) - 2(\dot{A}+\dot{B})\right\}h^2(z)\right]$$

$$= \alpha_{10}, \quad R_{0303} = -\frac{1}{4}\left(\frac{2C\ddot{C} - \dot{C}^2}{C}\right) = \alpha_3, \quad R_{0312} = -\frac{1}{4}(A+B)\left(\frac{\dot{A}B - A\dot{B}}{AB}\right) = \alpha_{11},$$

$$R_{1212} = \frac{1}{4ABC}\left[(A+B)^2 + \dot{A}\dot{B}C\right]\left[Af^2(z) + Bh^2(z)\right]\left[Ah^2(z) + Bf^2(z)\right] = \alpha_4,$$

$$R_{1313} = \frac{1}{4AB}\left[\{A^3 - AB(2A+3B-\dot{A}\dot{C})\}f^2(z) - \{AB(3A+2B-\dot{B}\dot{C}) - B^3\}h^2(z)\right] = \alpha_5,$$

$$R_{1323} = \frac{1}{4AB}\left[A^2(A-5B) + B^2(B-5A) + AB\dot{C}(\dot{A}+\dot{B})\right]f(z)h(z) = \alpha_{12},$$

$$R_{2323} = \frac{1}{4AB}\left[\{A^3 - AB(2A+3B-\dot{A}\dot{C})\}h^2(z) - \{AB(3A+2B-\dot{B}\dot{C}) - B^3\}f^2(z)\right] = \alpha_6.$$

Writing the curvature tensor with components $R_{abcd}$ at $p$ as a $6 \times 6$ symmetric matrix

$$R_{abcd} = \begin{pmatrix} \alpha_1 & \alpha_7 & 0 & 0 & \alpha_8 & \alpha_9 \\ \alpha_7 & \alpha_2 & 0 & 0 & \alpha_{10} & \alpha_8 \\ 0 & 0 & \alpha_3 & \alpha_{11} & 0 & 0 \\ 0 & 0 & \alpha_{11} & \alpha_4 & 0 & 0 \\ \alpha_8 & \alpha_{10} & 0 & 0 & \alpha_5 & \alpha_{12} \\ \alpha_9 & \alpha_8 & 0 & 0 & \alpha_{12} & \alpha_6 \end{pmatrix}. \tag{9}$$

It is important to note that we will consider the Riemann tensor components as $R^a{}_{bcd}$ for calculate CCS. Since we know from theorem [4] that when the rank of the $6\times 6$ Riemann matrix is greater than three there exists no proper CC. Hence we will consider only those cases where the rank of the $6\times 6$ Riemann matrix is less than or equal to three. There exist six, fifteen and twenty cases when the rank of the $6\times 6$ Riemann matrix is one, two and three respectively, i.e. we have altogether forty one cases when the rank of the $6\times 6$ Riemann matrix is less than or equal to three where proper CC may exist. Suppose the rank of the above $6\times 6$ Riemann matrix (9) is one. Then there is only one non zero row or column in (9). If we set five rows or columns are identically zero in (9) then there exist six possibilities when the rank of the $6\times 6$ Riemann matrix is one. In these six possibilities five give the contradiction and only one will arise which is given in case (II). For example consider the case when the rank of the $6\times 6$ Riemann matrix is one i.e. $\alpha_2 = \alpha_3 = \alpha_6 = \alpha_7 = \alpha_8 = \alpha_9 = \alpha_{10} = \alpha_{11} = \alpha_{12} = 0$ and $\alpha_1 \neq 0$. The constraints $\alpha_2 = 0 \Rightarrow \alpha_1 = 0$ which gives contradiction (here we assume that $\alpha_1 \neq 0$). Hence this is not possible. Now suppose rank of the above $6\times 6$ Riemann matrix (9) is two. Then there is only two non zero row or column in (9). If we set four rows or columns are identically zero in (9) then there exist fifteen possibilities when the rank of the $6\times 6$ Riemann matrix is two. If one proceed further after some lengthy calculation one finds that non of the above fifteen cases for rank two exists. By similar analysis we come to the conclusion that there exist only two cases when the rank of the $6\times 6$ Riemann matrix is three or less which are

(I) Rank=3, $\alpha_1 = \alpha_2 = \alpha_3 = \alpha_7 = \alpha_8 = \alpha_9 = \alpha_{10} = \alpha_{11} = 0$, $\alpha_4 \neq 0$, $\alpha_5 \neq 0$, $\alpha_6 \neq 0$ and $\alpha_{12} \neq 0$.

(II) Rank=1, $\alpha_4 \neq 0$, and $\alpha_1 = \alpha_2 = \alpha_3 = \alpha_5 = \alpha_6 = \alpha_7 = \alpha_8 = \alpha_9 = \alpha_{10} = \alpha_{11} = \alpha_{12} = 0$.

We will consider each case in turn.

**Case (I):**

In this case we have $\alpha_4 \neq 0$, $\alpha_5 \neq 0$, $\alpha_6 \neq 0$, $\alpha_{12} \neq 0$, $\alpha_1 = \alpha_2 = \alpha_3 = \alpha_7 = \alpha_8 = \alpha_9 = \alpha_{10} = \alpha_{11} = 0$, the rank of the $6\times 6$ Riemann matrix is three and there exists a unique (up to a multiple) no where zero time like vector field $t_a = t_{,a}$ solution of equation (3) and $t_{a;b} \neq 0$. From the above constraints we have $B(t) = d\, A(t)$, $C(t) = e A(t)$ and $A(t) = (at+b)^2$, where $a,b,d,e \in R (d,e > 0)$. The line element in this case takes the form

$$ds^2 = -dt^2 + (at+b)^2 \left[ \begin{array}{c} (f^2(z)+eh^2(z))dx^2 + (dh^2(z)+f^2(z))dy^2 + \\ 2(1+d)f(z)h(z)dx\,dy + e\,dz^2 \end{array} \right]. \tag{10}$$

Substituting the above information into the CCS equations one find that [22]

$$X^0_{,1} = X^0_{,2} = X^0_{,3} = 0,\ X^1_{,0} = X^1_{,3} = 0,\ X^2_{,0} = X^2_{,3} = 0,\ X^3_{,0} = X^3_{,1} = X^3_{,2} = 0, \tag{11}$$

$$R^1{}_{212,3}X^3 + 2R^1{}_{112}X^1_{,2} + R^1{}_{212}X^2_{,2} = 0, \tag{12}$$

$$R^2{}_{112,3}X^3 + 2R^2{}_{112}X^1_{,1} - 2R^1{}_{112}X^2_{,1} = 0, \tag{13}$$

$$R^1{}_{112,3}X^3 + R^1{}_{112}\left(X^1_{,1} + X^2_{,2}\right) + R^1{}_{212}X^2_{,1} + R^2{}_{112}X^1_{,2} = 0, \tag{14}$$

$$R^1{}_{313,3}X^3 + 2R^1{}_{313}X^3_{,3} + R^1{}_{323}\left(X^2_{,1} + X^1_{,2}\right) = 0, \tag{15}$$

$$R^1{}_{323,3}X^3 + 2R^1{}_{323}X^3_{,3} + \left(R^1{}_{313} - R^1{}_{323}\right)X^1_{,1} + R^1{}_{323}X^2_{,2} - R^2{}_{323}X^1_{,2} = 0, \tag{16}$$

$$R^2{}_{323,3}X^3 + 2R^2{}_{323}X^3_{,3} + R^2{}_{313}X^1_{,2} - R^1{}_{323}X^2_{,1} = 0. \tag{17}$$

Equation (11) gives $X^0 = K(t), X^1 = F^1(x,y), X^2 = F^2(x,y)$ and $X^3 = F^3(z)$, where $K(t)$ is an arbitrary function of $t$ only and $F^1(x,y), F^2(x,y)$ and $F^3(z)$ are functions of integration. If one proceeds further one find that curvature collineations in this case [22] are

$$X^0 = K(t),\ X^1 = -c_1 y + c_2,\ X^2 = -c_1 x + c_3,\ X^3 = c_1, \tag{18}$$

where $c_1, c_2, c_3 \in R$. One can write the above equation (18) subtracting Killing vector fields as

$$X = (K(t),0,0,0). \tag{19}$$

Proper curvature collineation in this case clearly form an infinite dimensional vector space. It is important to note that the constants $a$ and $b$ can not be zero simultaneously.

**Case (II):**

In this case we have $\alpha_1 = \alpha_2 = \alpha_3 = \alpha_5 = \alpha_6 = \alpha_7 = \alpha_8 = \alpha_9 = \alpha_{10} = \alpha_{11} = \alpha_{12} = 0,\ \alpha_4 \neq 0$ and the rank of the $6 \times 6$ Riemann matrix is one. Here, there exist two linear independent solutions $t_a = t_{,a}$ and $z_a = z_{,a}$ of equation (3). The vector field $t_a$ is covariantly constant whereas $z_a$ is not covariantly constant. From the above constraints we have $A(t) = B(t) = C(t) = (t+q)^2$, where $q \in R$. The line element takes the form

$$ds^2 = -dt^2 + (t+q)^2 \left[ \begin{array}{c} (f^2(z)+h^2(z))dx^2 + (h^2(z)+f^2(z))dy^2 + \\ 4f(z)h(z)dx\,dy + dz^2 \end{array} \right]. \tag{20}$$

Curvature collineations in this case [22]

$$X^0 = N(t,z),\ X^1 = -c_1 y + c_2,\ X^2 = -c_1 x + c_3,\ X^3 = c_1, \tag{21}$$

where $c_1, c_2, c_3 \in R$ and $N(t,z)$ is an arbitrary function of $t$ and $z$. One can write the above equation (21) subtracting Killing vector fields as

$$X = (N(t,z), 0, 0, 0). \tag{22}$$

Proper curvature collineation in this case clearly form an infinite dimensional vector space.

## SUMMARY

In this paper an attempt is made to explore all the possibilities when the Bianchi types $VI_0$ and $VII_0$ space-times admit proper CCS. An approach is adopted to study proper CCS of the above space-times by using the rank of the $6 \times 6$ Riemann matrix and also using the theorem given in [4], which suggested where proper curvature collineations exist. From the above study we obtain the following results:

(i) We obtain the space-time (10) that admits proper curvature collineations when the rank of the $6 \times 6$ Riemann matrix is three and there exists a unique nowhere zero independent timelike vector field, which is the solution of equation (3) and is not covariantly constant. In this case proper curvature collineations form an infinite dimensional vector space (for details see case I).

(ii) The space-time (20) is obtained, which admits proper curvature collineations (see case II) when the rank of the $6 \times 6$ Riemann matrix is one and there exist two independent solutions of equation (3) but only one independent covariantly constant vector field. In this case proper curvature collineations form an infinite dimensional vector space.

## References

[1] G. H. Katzin, J. Levine and W. R. Davis, J. Math. Phys. **10** (1969) 617.
[2] G. Shabbir, Amjad Ali and M. Ramzan, Gravitation and Cosmology, **16** (2010) 61.
[3] G. Shabbir and Amjad Ali, Adv. Studies Theor. Phys., **4** (2010) 151.
[4] G. S. Hall and J. da. Costa, J. Math. Phys. **32** (1991) 2854.
[5] G. S. Hall, Symmetries and Curvature Structure in General Relativity, World Scientific, 2004.

[6] G. Shabbir, Gravitation and Cosmology, **9** (2003) 139.

[7] G. Shabbir, Nuovo Cimento B, **118** (2003) 41.

[8] G. S. Hall and G. Shabbir, Classical Quantum Gravity, **18** (2001) 907.

[9] G. S. Hall, Gen. Rel. Grav. **15** (1983) 581.

[10] A. H. Bokhari, A. Qadir, M. S. Ahmed and M. Asghar, J. Math. Phys. **38** (1997) 3639.

[11] R. A. Tello-Llanos, Gen. Rel. Grav. **20** (1988) 765.

[12] J. Carot and J. da. Costa, Gen. Rel. Grav. **23** (1991) 1057.

[13] G. S. Hall, Classical Quantum Gravity, **23** (2006) 1485.

[14] G. S. Hall and Lucy MacNay, Classical Quantum Gravity, **22** (2005) 5191.

[15] A. H. Bokhari, M. Asghar, M. S. Ahmed, K. Rashid and G. Shabbir, Nuovo Cimento B, **113** (1998) 349.

[16] G. Shabbir, A. H. Bokhari and A. R. Kashif, Nuovo Cimento B, **118** (2003) 873.

[17] G. S. Hall and G. Shabbir, Gravitation & Cosmology, **9** (2003) 134.

[18] G. Shabbir, Nuovo Cimento B, **119** (2004) 433.

[19] G. Shabbir, Nuovo Cimento B, **121** (2006) 319.

[20] G. Shabbir and Abu Bakar Mehmood, Modern Physics Letters A, **22** (2007) 807.

[21] G. Shabbir and M. Ramzan, International journal of Modern Physics A, **23** (2008) 749.

[22] Amjad Ali, Ph. D. Thesis, GIK Institute, Pakistan, 2009.

[23] H. Stephani, D. Kramer, M. A. H. MacCallum, C. Hoenselears and E. Herlt, Exact Solutions of Einstein's Field Equations, Cambridge University Press, 2003.